# A Comprehensive Model to achieve Service Reusability for Multi level stakeholders using Non-Functional attributes of Service Oriented Architecture

Shanmugasundaram .G #1, V. Prasanna Venkatesan#2, C.Punitha Devi *3

\# Department of Banking Technology, * Department of Computer Science & Engg., Pondicherry University
R.V. Nagar, Kalapet, Puducherry-605004, India

[1] sundar_gss2004@yahoo.co.in
[2] prasanna_v@yahoo.com
[3] c.punithadevi@gmail.com

*Abstract*— SOA is a prominent paradigm for accomplishing reuse of services. Service reusability is one dominant factor which has a greater influence on achieving quality in SOA systems. There exists sufficient research in this area and researchers have contributed many works towards achieving quality in SOA systems but much emphasis was not provided on service reusability [1] [2] [3]. Few authors have addressed reusability factor with limited non- functional attributes. Our study focuses on identifying the non-functional attributes which have major or greater influence towards obtaining reusability in SOA systems. The objective of this study goes into the next level, to categorize the non-functional attributes on multi stakeholder's perspective i.e. Service Consumer, Service Provider and Service Developer which paves the way to build a comprehensive quality model for achieving Service Reusability.

## I. INTRODUCTION

SOA acts as the major platform for building distributed applications which cross organizational boundaries because of its flexible, heterogeneous and loosely coupled nature. SOA-based businesses applications can span several networked enterprises, with services that encapsulate and externalize various corporate applications and data collections. Popularity of SOA applications not only based on its functionalities also delivers in high quality. On defining quality attributes for SOA, Service reusability stands as one of the essential factor. Reusability was the core of SOA and has helped in gaining its popularity. Service reusability is the key determinant factor for identification of optimally granular services since it proves its role in saving cost of development, and maintenance.

Achieving reusability is not a simple task as it has influence on different stakeholders. Addressing of this feature leads to two main issues or questions

- What are all the attributes (functional and non-functional) having greater impact on reusability principle?
- Are they any measures or model is available to achieve it completely.

Our objective focuses on identifying the non-functional attributes that has high impact on reusability. Later the identified attributes need to be categorized for different stakeholders. Hence this work gives a complete picture of reusability factors or attributes and their categorization on stakeholder's perspective.

The rest of paper is organized as follows; section 2 gives the review of related work on SOA quality and Service Reusability. Section 3 elaborates outcome of the reviews listed in section 2. Section 4 delivers our comprehensive quality model. Finally section 5 gives conclusions and future directions towards service reusability.

## II. RELATED WORKS

Quality attributes

Quality attributes are essential in choosing and designing an architecture style of any system. In SOA, quality attributes inherently affect the business goal as defined quality attributes have a greater impact on business decisions. The survey initially starts with an objective to list works in qualities of SOA and second part of the review covers the works related to service reusability factor of SOA.

Defining the quality attributes for SOA raises these questions
1. What are the quality attributes that have an impact on business goal of SOA systems?
2. Which category does the quality attributes fits in?
3. Are there any measure or evaluation mechanism to check whether the attributes are properly addressed?
The related work has been reviewed considering the above questions

**Survey about quality attributes**
- [Balfagih and Hassan 2009] examined the various quality attributes of SOA and Web Services and classified them into different perspectives i.e. developer, provider and consumer.
- [Glaster et al.] Identified the critical importance and the difficulties associated with handling Non-Functional parameters in general and the fact that they are even more difficult to address in the SOA context. They made an attempt to generate a checklist of NFPs for SOA to be used by the service providers.
- [Choi et al. 2008] have identified some of the unique features of SOA and then derived six quality attributes and proposed the corresponding metrics to measure each quality attribute.
- [Liam O'Brien Lero et al. 2007] have discussed the SOA aspects related to various quality attributes

**Review towards quality attributes for service reusability**
The review listed below describes the existing works on service reusability. Here the review indicates that contributions and work of researchers is towards defining the functional and non-functional attributes along with the metrics for measuring reusability. Some have addressed specifically the functional attributes like service cohesion, coupling and granularity. Few authors have listed reusability as the key quality attribute to Service Developer.

- [Si Won Choi and Soo Dong Kim, 2008] proposed a comprehensive quality model for evaluating reusability with functional attributes of modularity and commonality and non - functional attributes of discoverability and availability.

- [Renuka sindhgatta, et al., 2009] addressed the functional attributes like coupling and cohesion and non-functional attributes like composability and reusability, dervied metrics for coupling and cohesion and also for composability and reusability

- [Zain Balfagih and Mohd Fadzil, 2009] listed the qualities based on the different stakeholders, they addresssed reusability as a major quality for service developer.

- [George Feuerlicht 2011] stated that service granularity has impact on service reusability.

- Perepletchikov, et al., 2007 discusses about service coupling and cohesion which support reusability factor. They have defined metrics for the two functional attributes.

- [Mikhail Perepletchikov, et al., 2010] gives impact of service cohesion and coupling on reusability

- [Saad Alahmari, et al., 2011] defined metrics for service granularity to achieve reusability.

From this review we could conclude that NFA relating to service reusability are not addressed completely and which has motivated us to carry out the problem.

III. OUR CONTRIBUTION

The objective of our proposed is to identify the non-functional attributes of Service Reusability and to categorize it to multilevel stakeholders. Reusability in SOA cannot be addressed as a feature that satisfies some properties nor is it a separate entity. The Attributes of reusability has direct or indirect impact. To ensure service reusability we need to address all the NFA which have positive or negative impact.

To achieve complete reusability in SOA systems we need to define the factors for different stakeholder's perspectives. From the related work the attributes that have influence on reusability could be identified.
For example if discovery of services is easier then the reusability of services will be more hence service discovery shows positive impact.
Likewise other NFA's could also be related

TABLE I
LIST OF NFA'S RELATES TO SERVICE REUSABILITY

| S. No. | Non-Functional Attributes (NFA's) | Researchers |
|---|---|---|
| 1. | Usability | [Zain Balfagih and Mohd Fadzil Hassan ,2009][ Si Won Choi, Jin Sun Her, and Soo Dong Kim 2007] |
| . | Effectiveness | [Bingu Shim, et al., 2008] [Zain Balfagih and Mohd Fadzil Hassan ,2009] |
| 3. | Conformance | [D. J. Artus, 2006] [Zain Balfagih and Mohd Fadzil Hassan |
|     |               | ,2009] [Si Won Choi, Jin Sun Her, and Soo Dong Kim 2008]                                                           |
| 4.  | Testability   | [Liam O'Brien, et al., 2007]                                                                                       |
| 5.  | Composability | [Zain Balfagih and Mohd Fadzil Hassan ,2009] [Si Won Choi and Soo Dong Kim, 2009] [T. Erl. 2006] [T. Erl. 2009]    |
| 6.  | Availability  | [Si Won Choi, Jin Sun Her, and Soo Dong Kim 2007]                                                                  |
| 7.  | Adaptability  | [Si Won Choi, Jin Sun Her, and Soo Dong Kim 2007]                                                                  |
| 8.  | Reliability   | [Si Won Choi, Jin Sun Her, and Soo Dong Kim 2007]                                                                  |
| 9.  | Flexibility   | [Bingu Shim, et al., 2008]                                                                                         |
| 10. | Discoverability | [Si Won Choi and Soo Dong Kim, 2009] [Si Won Choi, Jin Sun                                                       |
|     |               | Her, and Soo Dong Kim 2007] [Zain Balfagih and Mohd Fadzil Hassan ,2009] [T. Erl. 2006] [T. Erl. 2009]             |
| 12. | Modifiability | [Zain Balfagih and Mohd Fadzil Hassan ,2009]                                                                       |
| 13. | Security      | [Zain Balfagih and Mohd Fadzil Hassan ,2009]                                                                       |

The second objective is to categorize the non-functional attributes and fit it into the various stakeholders. The stakeholders are in different forms, service provider, service consumer and service developer. Service developer, are one who originally develop or create the service, provider are the parties or organization who offers the service for consumption and finally service consumer are the party or enterprise who is to consume the services for their enterprise application / for developing new applications.

IV. COMPREHENSIVE MODEL FOR REUSABILITY OF SOA FOR MULTI-STAKEHOLDER'S USING NFA

The table 1 shows complete list of non-functional attributes. Based on the different researcher's contribution we can categorize the NFA for different stakeholders. The NFA's listed in Service developer can be in service consumer and also in service provider, similarly for other categories. Some of the attributes are common which falls in all categories that can have greater impact of reusability when compared to other attributes. Let us consider the attribute service discovery, it falls both in service consumer and in service developer, here the attribute is common for two category but the features ensuring discovery at consumer's perspective would be different from that of developer's perspective.

The comprehensive model (figure 1) gives clear picture to reusability factors completely addressed in SOA systems. To estimate or evaluate reusability of SOA systems completely focus has to be at levels

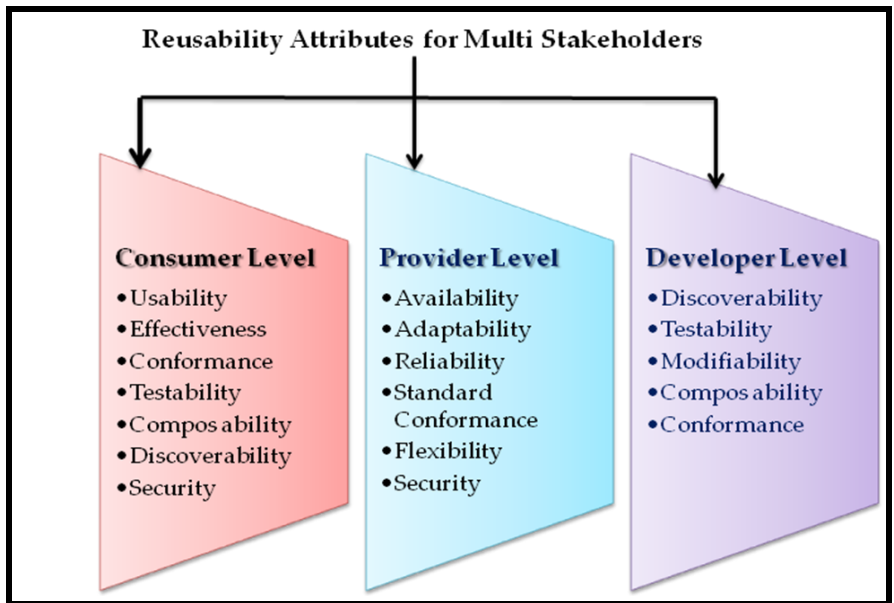

Figure 2 Comprehensive Model of Reusability based NFA for different stakeholders

V. DISCUSSION

The table below represents the various contributors towards the multi stakeholder perspective for service reusability using non-functional attributes. The notation * represents the indirect support and ^ represents the direct support. Most of the contributors address the influence of different stakeholders indirectly. They have not categorized the NFA for each stakeholder. The comparison of different works clearly states that NFA for Service reusability of various stakeholders was not addressed precisely. Our work lists the NFAs for multi stakeholders that would help to achieve service reusability completely.

TABLE III
COMPARISON OF VARIOUS WORKS WITH MULTI-STAKEHOLDER'S FOR SERVICE REUSABILITY

| Contributors | Multi stakeholder perspectives for Service Reusability Using NFA | | |
|---|---|---|---|
| | Service Consumer | Service Provider | Service Developer |
| Si Won Choi and Soo Dong Kim | * | * | * |
| Zain Balfagih and Mohd Fadzil | * | * | * |
| Glaster et al. | * | * | * |
| Renuka sindhgatta | | | * |
| Our Proposed Model | ^ | ^ | ^ |

VI. CONCLUSION

Different works has been done to discuss the qualities of SOA. Most of current efforts of SOA quality has not focused on reusability factor and also has not considered the multi stakeholders with the reusability. In this paper we have presented the comprehensive model for reusability that uses non-functional attributes for various stakeholders. The proposed model shows the way to achieve reusability in all levels thereby enabling complete reusability of the SOA systems. Our future work will be on proposing the measures for each NFA's of different level to completely evaluate the reusability of the entire SOA systems.